\documentclass[aps,pre,twocolumn,groupedaddress,showpacs]{revtex4}

\usepackage{graphicx}%

\begin{document}

\title{Single- and double-vortex vector solitons in self-focusing nonlinear media}

\author{Jos\'e R. Salgueiro and Yuri S. Kivshar}

\affiliation{Nonlinear Physics Centre, Research School of Physical
Sciences and Engineering, Australian National University, Canberra
ACT 0200, Australia}

\begin{abstract}
We study two-component spatial optical solitons carrying an
angular momentum and propagating in a self-focusing saturable
nonlinear medium. When one of the components is small, such vector
solitons can be viewed as a self-trapped vortex beam that guides
either the fundamental or first-order guided mode, and they are
classified as single- and double-vortex vector solitons. For such
composite vortex beams, we demonstrate that a large-amplitude
guided mode can stabilize the ring-like vortex beam which usually
decays due to azimuthal modulational instability. We identify
different types of these vector vortex solitons and demonstrate
both vortex bistability and mutual stabilization effect.
\end{abstract}

\pacs{42.65.Tg, 05.45.Yv, 47.20.Ky}

\maketitle

\section{Introduction}

Vortices are fundamental localized objects which appear in many
branches of physics~\cite{pismen}. In fluid mechanics, coherent
structures in the form of vorticity filaments are central
dynamical objects to understand most fluid flows and particularly
fluid turbulence. More recently, the study of vortices in
Bose-Einstein condensates revealed many intriguing properties of
superfluids created by ultra-cold atoms~\cite{BEC}. Different
types of vortices can also be found and identified in optics; one
of the simplest objects of this kind is {\em a phase singularity}
in an optical wave front which is associated with a phase
dislocation carried by a diffracting optical
beam~\cite{review,list}.

In self-focusing saturable nonlinear media, optical vortices can
exist as self-trapped ring-like optical beams with zero intensity
at the center carrying a phase singularity~\cite{kruglov}.
However, due to the self-focusing nature of nonlinearity such
ring-like vortex beams become unstable to azimuthal perturbations,
and they decay into several fundamental optical solitons flying
off the main ring~\cite{firth}. This effect was observed
experimentally in different nonlinear media, including the
saturable Kerr-like nonlinear media, biased photorefractive
crystals, and quadratic nonlinear crystals operating in the
self-focusing regime (see details and references in
Ref.~\cite{book}). There are known several ways to stabilize this
azimuthal modulational instability, including the vortex
stabilization in the presence of a large-amplitude beam guided by
it~\cite{dipole2}, and the stabilizating effect of partial
incoherence of light on the vortex~\cite{our_prl}.

When a self-trapped vortex beam guides a large-amplitude
fundamental beam,  it creates together with the guided beam a
composite object in the form of {\em  a vector
soliton}~\cite{book}. Mutual coupling between the fundamental beam
and the vortex-carrying beam can create different novel types of
composite vector solitons carrying an angular
momentum~\cite{vortex0,vortex,vortex1}. The properties of such
vector vortex solitons can differ substantially from the
properties of one-component scalar vortices and scalar solitons.
In particular, the mutual coupling between the beams can modify
dramatically the vortex properties and, in particular, can
suppress the development of azimuthal instability~\cite{dipole2}.

In this paper, we analyze the existence, general properties, and
stability of the vector vortex solitons in a self-focusing
saturable nonlinear medium. We study two types of such
two-component composite vortex solitons, i.e. {\em single-vortex
vector solitons} that can be considered as the vortex-induced
waveguide that guides a fundamental mode, and  {\em double-vortex
vector solitons}, when the localized field is similar to the
first-order guided mode, being a vortex beam by itself. For some
of the cases, we demonstrate that a mutual incoherent coupling
between the vortex waveguide and a large-amplitude guided mode it
guides can provide a strong stabilizing mechanism for stable or
quasi-stable two-component vortex solitons to exist in such media,
in agreement with the recent observation~\cite{china}  of the
stabilizing mechanism of the mutual coupling between different
components of the composite vortex beam.

\section{Model}

In  order to study the vector vortex solitons, we consider the
interaction of two mutually incoherent optical beams propagating
in a self-focusing nonlinear saturable medium. The evolution
equations for two incoherently interacting beams can be presented
in the following dimensionless form,
\begin{equation}
\label{vortex_induced_vaweguides}
   \begin{array}{l} {\displaystyle
      i \frac{\partial u}{\partial z} +\Delta_{\perp} u +
        \frac{(|u|^2 + \mu |v|^2) u}{1+\sigma (|u|^2+|v|^2)} = 0,
    } \\*[9pt] {\displaystyle
      i \frac{\partial v}{\partial z} +\Delta_{\perp} v +
        \frac{(|v|^2 + \mu |u|^2) v}{1+\sigma (|u|^2+|v|^2)} = 0,
      }
   \end{array}
\end{equation}
where $u$ and $v$ are the dimensionless amplitudes of the fields,
the parameter $\sigma$ characterizes the nonlinearity saturation
effect, and the mutually incoherent interaction between the modes
is described by the coupling parameter $\mu$. The spatial
coordinate $z$  is the beam propagation direction, and
$\Delta_\perp$ stands for the transversal part of the Laplace
operator in the cylindrical coordinates, defined through the
radial, $r=(x^2+y^2)^{1/2}$, and angular, $\phi =\tan^{-1}(y/x)$,
coordinates. Model (\ref{vortex_induced_vaweguides}) provides a
straightforward generalization to a number of important cases
studied earlier. In particular, the limit $\sigma \rightarrow 0$
corresponds to the Kerr medium with cubic
nonlinearity~\cite{vortex1} where all self-trapped beams may
undergo collapse instability. The case of the saturable
nonlinearity at $\mu=1$ corresponds to the incoherent beam
interaction in photorefractive nonlinear
media~\cite{vortex0,vortex,dipole3}.
We look for stationary solutions of the system
(\ref{vortex_induced_vaweguides}) that describe a radially
symmetric single-charged vortex beam in the field $u$,
\begin{equation}
\label{eq_u} u (r, \phi; z)=u(r)e^{i\phi}e^{iz},
\end{equation}
where the amplitude $u(r)$ vanishes for $r\rightarrow \infty$. We
assume that the vortex (\ref{eq_u}) guides (or is coupled to) the
second beam,
\begin{equation}
\label{eq_v}
 v(r, \phi; z)=v(r)e^{il\phi}e^{i\beta z},
 \end{equation}
where $\beta$ is a dimensionless ratio of the propagation
constants. In Eqs.~(\ref{eq_u}), (\ref{eq_v}) the functions $u(r)$
and $v(r)$ are the  radial envelopes of the interacting fields,
and $l$ ($l=0, \pm 1$) is the angular momentum of the guided mode.
Equations for the stationary envelopes are given by
\begin{equation}
\label{stationary_vortex_induced_vaweguides}
   \begin{array}{l} {\displaystyle
      - u +\Delta_r u -\frac{1}{r^2}u +
        \frac{(u^2 + \mu v^2) u}{1+\sigma (u^2+v^2)} = 0,
    } \\*[9pt] {\displaystyle
      -\beta v +\Delta_r v -\frac{l^2}{r^2}v +
        \frac{(v^2 + \mu u^2) v}{1+\sigma (u^2+v^2)} = 0,
      }
   \end{array}
\end{equation}
where $\Delta_r$ is the radial part of the Laplace operator,
\[
\Delta_r \equiv \frac{1}{r} \frac{d}{dr} \left(r~\frac{d}{dr}
\right).
\]
The radially symmetric, spatially localized solutions of the
system (\ref{stationary_vortex_induced_vaweguides}) describe
different types of two-component composite solitons carrying an
angular momentum, and they can form either single- or
double-vortex vector solitons. In a two-dimensional geometry, such
solutions can only be found numerically.

\section{Single-vortex vector solitons}
\label{vortex_fundamental_mode}

We consider the self-trapped vortex beam created in the main field
$u$, with the asymptotic behavior $u(r) \rightarrow 0$ as $r
\rightarrow \infty$. Besides that, we require that  the
incoherently interacting component $v(r)$ describes a localized
mode, i.e. $v(r) \rightarrow 0$ and therefore $\beta>0$. At the
origin ($r=0$), the boundary condition for the vortex is $u=0$
and, if we seek the single-vortex solution with the fundamental
guided mode ($l=0$), the corresponding boundary condition for
$v(r)$ is $dv/dr=0$ (or $v=v_0$).

\begin{figure}
\centerline{\includegraphics[width=3.2 in]{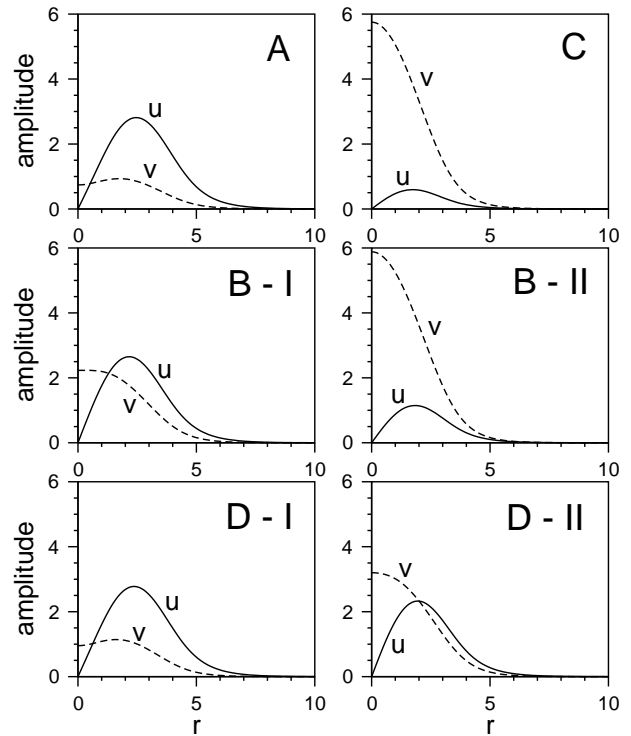}}
\caption{\label{fig5}
Examples of single-vortex vector solitons created by the
vortex beam $u$ and the fundamental mode $v$ it guides. Labels
correspond to the points marked in Fig.~\ref{fig4}. For points B
and D, we show two different solutions named with roman numbers I
and II, that exist due to bistability.  The model parameters are:
A $(\mu=1.10, \beta=1.38)$, B $(\mu=1.10, \beta=1.44)$, C
$(\mu=1.17, \beta=1.38)$, and D $(\mu=1.17, \beta=1.47)$. }

\end{figure}

\begin{figure}
\centerline{\includegraphics[width=3.2 in]{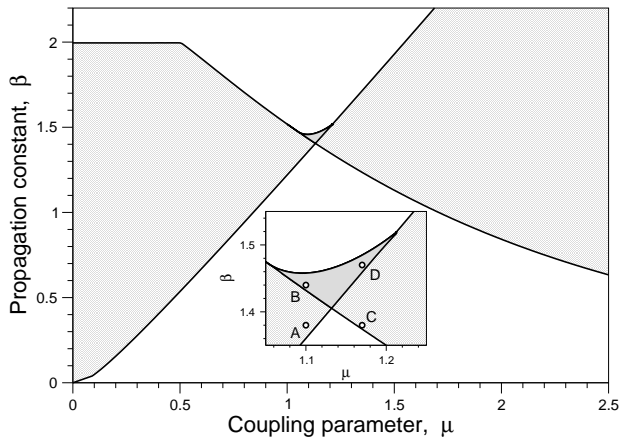}}
\caption{\label{fig4}
Existence domain of the single-vortex vector solitons on
the parameter plane ($\beta$, $\mu$) at $\sigma=0.5$. Inset shows
an enlarged region of bistability where two different types of
vortex-mode solutions coexist. }
\end{figure}

\begin{figure}
\centerline{\includegraphics[width=3.2 in]{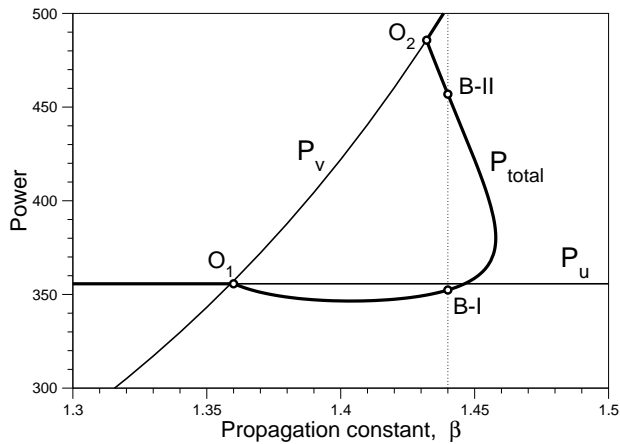}}
\caption{\label{fig6}
Bifurcation diagram for the two-component vector vortex
solitons, shown together with the families of the scalar
fundamental and vortex solitons at $\mu=1.1$. Thin curves
represent the powers $P_u$ and $P_v$ of each of the scalar
solitons; the thick curve is the total power $P_{\rm total}$ of a
composite vortex soliton. Points O$_1$ and O$_2$ are the
bifurcation points where the vector soliton emerges, and the
intersection points B-I and B-II correspond to the example of
bistable single-vortex solutions for the point B ($\beta=1.44$)
shown in Fig.~\ref{fig5}.} 
\end{figure}

We find localized solutions numerically, by means of the
relaxation technique. In Fig.~\ref{fig5}, we show some examples of
two-component localized solutions which describe a fundamental (no
nodes) beam, guided by the self-trapped vortex that create
together {\em a single-vortex vector soliton}. The existence
domain for such solutions has been calculated numerically for a
special case $\sigma=0.5$, and it is shown in Fig.~\ref{fig4} on
the parameter plane ($\beta$, $\mu$). The existence region for
single-vortex vector solitons is composed of two regions which
overlap in a triangular-shaped domain shaded with a different
intensity in Fig.~\ref{fig5}. In this intersection, two types of
vortex solutions mark the familiar bistability phenomenon. The
existence domain is restricted by the solid curves which describe
some specific cutoff boundaries. Close to the lower cutoff, one of
the components becomes small: the fundamental field in the left
region (see the cases A and D-I), and the vortex in the right
region (cases B-II and C). The other component is therefore only
weakly distorted. However, when the propagation constant $\beta$
is close to the upper cutoff (cases B-I and D-II), the amplitudes
of both components become comparable, affecting strongly each
other. For small values of the coupling parameter $\mu$, the
existence region is limited from above by the constant value
$1/\sigma$ (in our example, $1/\sigma=2$), the value which can be
easily explained by a simple qualitative analysis. Indeed,
according to Eqs.~(\ref{stationary_vortex_induced_vaweguides}),
the existence of bounded stationary states for both $u$ and $v$
components requires the following conditions to be valid,
\begin{eqnarray}
\label{bounded_states_1}
\max{\left\{\frac{u^2+\mu v^2}{1+\sigma (u^2+v^2)}\right\}}>1,\\
\label{bounded_states_2} \max{\left\{\frac{v^2+\mu u^2}{1+\sigma
(u^2+v^2)}\right\}}>\beta.
\end{eqnarray}
Considering $\mu \rightarrow 0$, we obtain from
Eq.~(\ref{bounded_states_2}) that the existence of localized
solution requires that
\begin{equation}
\label{cond}
 \beta<\max \left\{\frac{v^2}{1+\sigma(u^2+v^2)}\right\}.
\end{equation}
The right-hand-side term vanishes when the amplitude of the
component $v$ vanishes at the cutoff, but it approaches the value
$1/\sigma$ when the amplitude of the field $v$ becomes large. This
analysis is valid for any type of the guided mode.

\begin{figure}
\centerline{\includegraphics[width=3.2in]{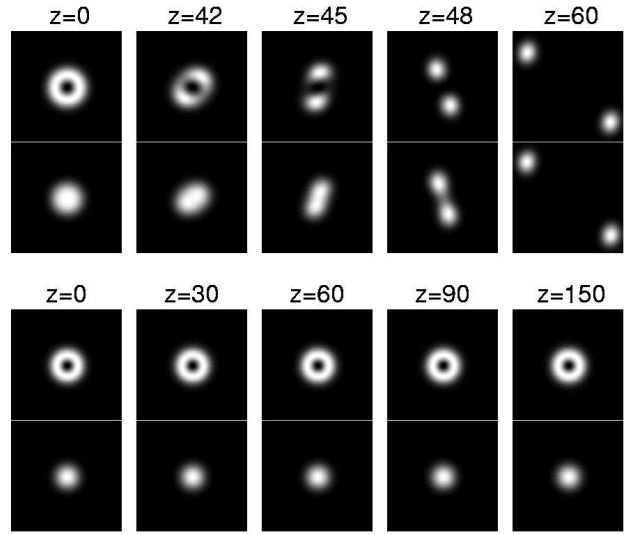}}
\caption{ \label{fig9}
Examples of the vortex propagation dynamics in the
bistability domain. Shown are the field intensity profiles as
gray-scale images at several propagation distances. Top: the
components of the B-I vector soliton. Bottom: the components of
the B-II vector soliton (see Fig.~1 and Fig.~3. ).}
\end{figure}

In order to describe the bistable vector solitons, in
Fig.~\ref{fig6} we display the bifurcation diagram of the
two-component vortex-mode localized solutions, for the partial and
total beam powers. Composite vortex-mode solitons presented by the
power dependence $P_{\rm total}$ originate at the bifurcation
point O$_1$ where the mode $v$ is small and can be described by
the linear theory. For larger value of $\beta$, this curve bends,
and then it merges with the other partial power curve $P_v$ at the
bifurcation point O$_2$ (see Fig.~\ref{fig6}). The bistable
solutions B-I and B-II, which are presented in Fig. 1 being
related to point B in Fig. 2, correspond to a single value of the
propagation constant $\beta$ in the bistability domain.
Importantly, two solutions have different stability properties,
and only one of them is stable, as shown in Fig.~\ref{fig9}. In
general, the solutions belonging to the left region in the domain
are unstable and those belonging to the right zone are stable.

\begin{figure} \centerline{\includegraphics[width=3.2
in]{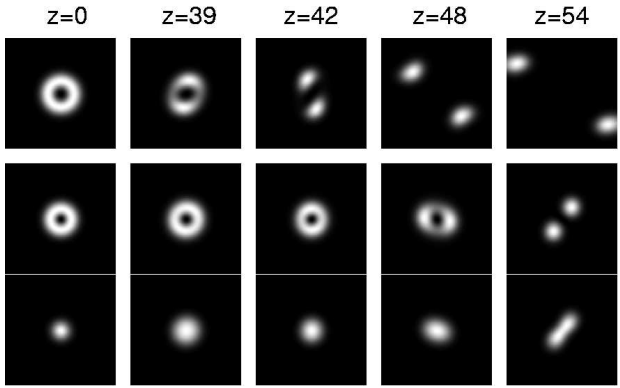}} \caption{\label{fig7}
Vortex stabilization due to mutual
interaction. Shown are the field intensity profiles as gray-scale
images at several propagation distances. Top: breakup of a scalar
vortex soliton ($v=0$). Bottom: both the vortex beam and the
fundamental guided mode (approximated by a Gaussian beam)
propagate together. Parameters are: $\sigma=0.5$, $\mu=1$, and
$\beta=1.45$.} 
\end{figure}

The incoherent interaction between the vortex beam and the
localized mode it guides has the character of attraction, and it
may provide an effective physical mechanism for stabilizing the
vortex beam in a self-focusing nonlinear medium. Indeed, a scalar
vortex beam becomes unstable in a self-focusing nonlinear medium
due to the azimuthal modulational instability. In this case, the
vortex splits into the fundamental beams that fly off the main
vortex ring. On the other hand, the bright solitons are known to
be stable in such media. We expect that a mutual attraction of the
components in a two-component beam may lead to a counter-balance
of the vortex instability by the bright component when its
amplitude is large enough. To confirm this idea, we consider a
two-component composite structure consisting of a vortex beam
together with the fundamental mode it guides, both described by
Eqs.~(\ref{vortex_induced_vaweguides}) at $\eta=1$ and $\mu=1.0$.
To study the mode stability, we propagate the stationary soliton
solutions numerically. In Fig.~\ref{fig7}, we compare the vortex
breakup for the scalar and vector systems. In the top row, we show
the propagation of a vortex alone in the scalar model; the vortex
breaks up into two solitons which fly away after some distance. In
the bottom row, we show the propagation of two coupled components
(the vortex and bright mode it guides). Due to a strong incoherent
coupling between the modes, the propagation of the vortex is
stabilized for some propagation distances,  so that the vortex
breakup is delayed dramatically, as shown in Fig.~\ref{fig7}(lower
row), or even become completely stable, similar to the case shown
in Fig.~\ref{fig9}(low row). We confirm this stabilization
mechanism by performing our study for a Gaussian input beam of the
bright component instead of the exact stationary state, as would
be easier realizing in experiment.

\section{Double-vortex vector solitons}

\begin{figure} \centerline{\includegraphics[width=3.2
in]{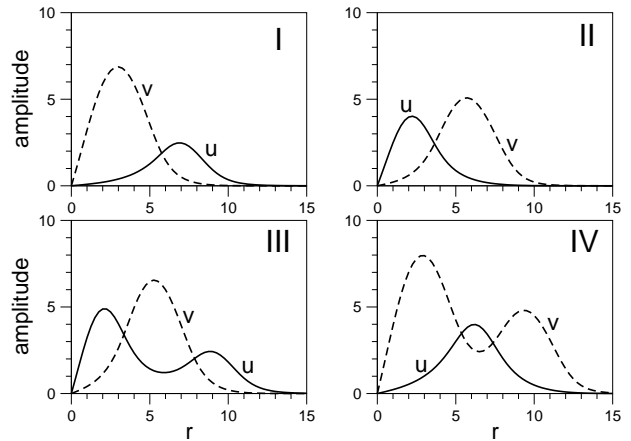}} \caption{\label{states_firstmode1}
Examples of four types of double-vortex
vector solitons calculated for $\sigma=0.5$, $\mu=0.3$ and
$\beta=1.5$. All solutions correspond to the point A in
Fig.~\ref{domain_first_mode}.} 
\end{figure}

\begin{figure} \centerline{\includegraphics[width=3.2
in]{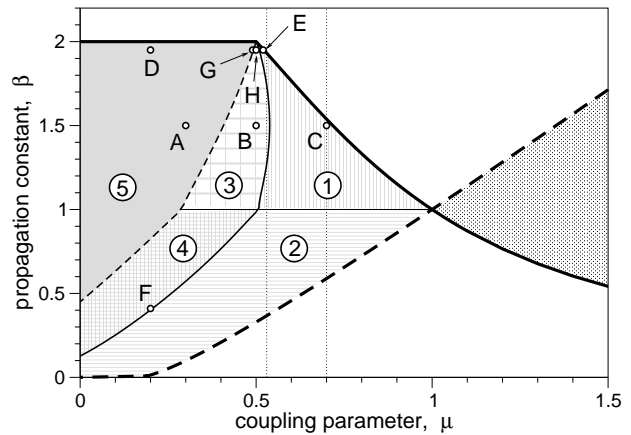}} \caption{\label{domain_first_mode}
Existence domain for different types of
double-vortex solitons. Domain 1: only the solutions of type I;
domain 2: only solution of type II; domain 3: solutions of types
I, II and III; domain 4: solutions of types I, II and IV; domain
5: all four types of solutions exist. Two vertical thin dotted
lines indicate the values of $\mu$ for which the bifurcation
diagrams are presented in Figs.~\ref{bifurcation_first_mode1} and
\ref{bifurcation_first_mode2}.}
\end{figure}

\begin{figure} \centerline{\includegraphics[width=3.2
in]{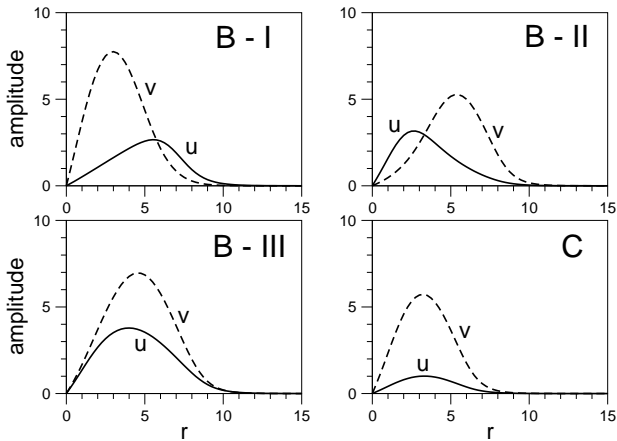}} \caption{\label{states_firstmode2}
Tristable solution found for $\mu=0.5$
and $\beta=1.5$ (point B in Fig.~\ref{domain_first_mode}) and the
unique solution found for $\mu=0.7$ and $\beta=1.5$ (point C). In
all cases $\sigma=0.5$.} 
\end{figure}

To study novel types of vector vortex solitons, we consider the
vortex beam in the field $u$ coupled to the first-order guided
mode described by the solution (\ref{eq_v}) with $l=1$. Thus, for
the field $v$ we look for a vortex-like localized solution with
the boundary condition $v(r)\rightarrow 0$ at $r=0$. Our analysis
shows that there exist four different kinds of such solutions, and
they form the families of the so-called {\em double-vortex vector
solitons}. We show some examples of these solutions in
Fig.~\ref{states_firstmode1} and Fig.~\ref{states_firstmode2},
whereas the existence domains for all solutions of this type are
shown on the parameter plane ($\beta$, $\mu$) in
Fig.~\ref{domain_first_mode}.

In the region labelled with number 5, there exist four types of
localized double-vortex solutions, as shown for the point A in
Fig.~\ref{states_firstmode1}. The first type (type I) is described
by two rings where the ring in the field $u$ is larger than the
ring in the filed $v$; the opposite situation occurs for the
solutions of type II (see Fig.~\ref{states_firstmode1}). Other two
types of solutions have one of the fields of a two-humped shape:
either the field $u$ (type III) or the field $v$ (type IV). As the
coupling parameter $\mu$ grows, the valley in the two-humped
solutions becomes shallower, and it disappears when the solution
crosses a thin dashed line on the existence plane (Fig.
~\ref{domain_first_mode}), moving either to the domains 3 or 4. In
the upper domain ($\beta>1$, domain 3), no solution of type IV can
be found further up from this line, being consequently the
boundary of its existence domain. Besides, the solution of type
III degenerates in a single-humped solution and still exists up to
the  thin continuous line (see Fig.~\ref{states_firstmode2}, case
B-III). The opposite occurs in the lower domain ($\beta<1$, domain
4) , where the solitons of type III cannot be found to the right
of the dashed line, and the solitons of type IV degenerate into
single-humped modes.

Single-humped solutions of types I and II both exist up to the
continuous thin line, though the maxima of both fields $u(r)$ and
$v(r)$ are shifted, approaching each other as the parameter $\mu$
increases (see the examples B-I and B-II in
Fig.~\ref{states_firstmode2}). Below this line, in the upper
domain ($\beta>1$, domain 3) the solutions II no longer exist,
while the solutions I have a symmetric shape where the maxima of
the fields approximately coincide (Fig.~\ref{states_firstmode2},
case C). Dashed thick curve corresponds to the cutoff for such
solutions, towards which the amplitude of the mode $v$ becomes
small and does not influence the vortex mode in the field $u$. On
the other hand, the continuous line is the upper cutoff where a
vector soliton originates.

\begin{figure} \centerline{\includegraphics[width=3.2
in]{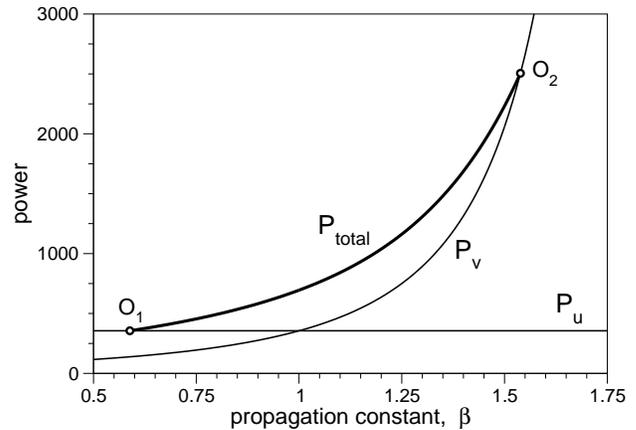}} \caption{\label{bifurcation_first_mode1}
Bifurcation diagram for vector and
scalar vortex solitons (at $\mu=0.7$). $P_u$ and $P_v$ (thin
lines) are the powers of scalar vortices created in each component
separately, while $P_{\rm total}$ (thick line) is the power of the
vector soliton originated at the bifurcation points O$_1$ and
O$_2$.} 
\end{figure}

\begin{figure} \centerline{\includegraphics[width=3.2
in]{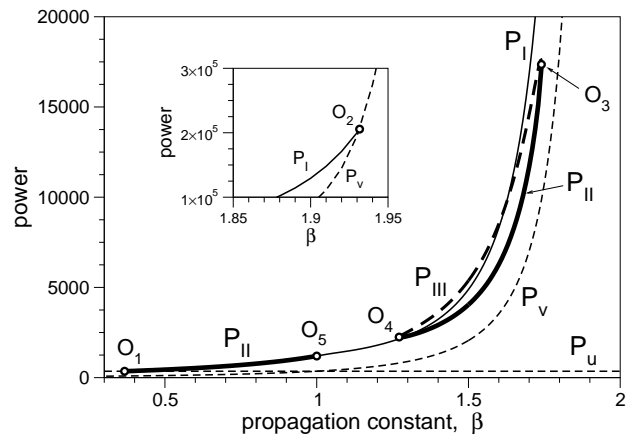}} \caption{Bifurcation diagram for vector and
scalar vortex solitons (at $\mu=0.53$). $P_u$ and $P_v$ (dashed
thin lines) are the powers of the scalar vortices created in each
component separately. Other curves are labelled with a subscript
corresponding to different types of vector solitons: type I
(continuous thin line), type II (continuous thick line), and type
III (dashed thick line). Points O$_1$ and O$_2$ are bifurcation
points, and O$_3$, O$_4$ and O$_5$ are critical points. An area
near the bifurcation point O$_2$ is shown in the inset.}
\label{bifurcation_first_mode2}
\end{figure}

\begin{figure} \centerline{\includegraphics[width=3.2
in]{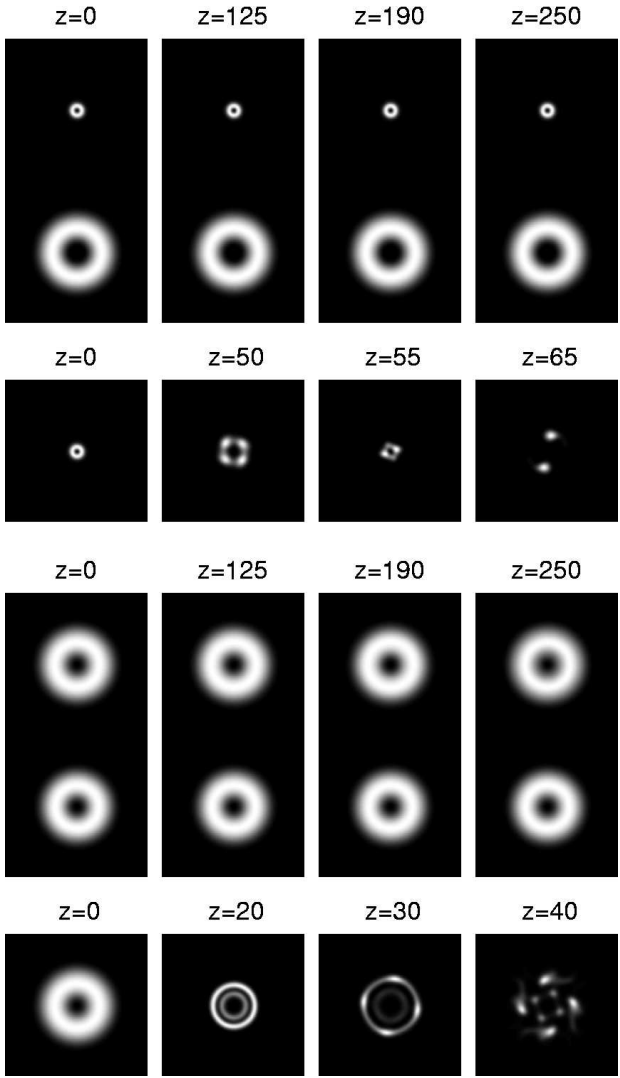}} \caption{\label{stable_prop}
Evolution of two types of stable
double-vortex vector solitons corresponding to point D (type II,
top rows) and point E (type I, bottom rows). For each soliton,
both components are shown for different values of the propagation
distance. For comparison, we show the corresponding unstable
evolution when the component $v$ is removed at the input.}
\end{figure}

\begin{figure} \centerline{\includegraphics[width=3.1
in]{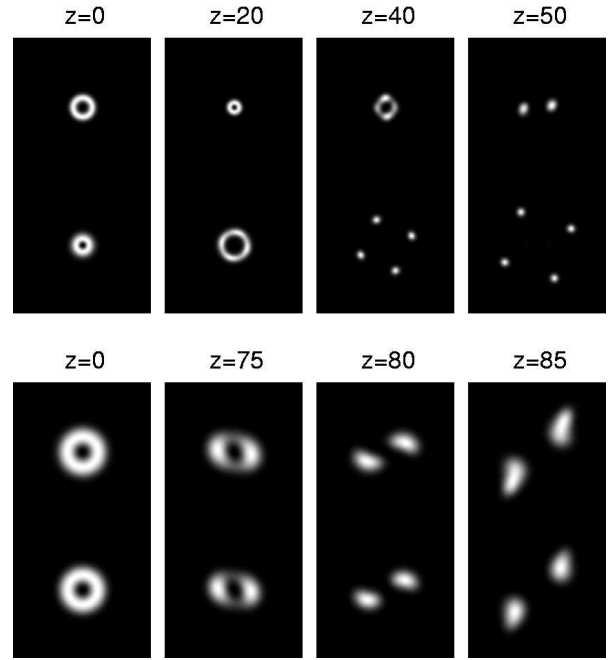}} \caption{Decay of the  vortex vector solitons into
2+4 (top row) and 2+2 (bottom row) fundamental solitons. The upper
case corresponds to the solution of type IV (point F; $\mu=0.2$,
$\beta=1.41$), and the lower case --to the solution of type I
(point C; $\mu=0.7$, $\beta=1.5$).} \label{decaying_states}
\end{figure}

\begin{figure} \centerline{\includegraphics[width=3.1
in]{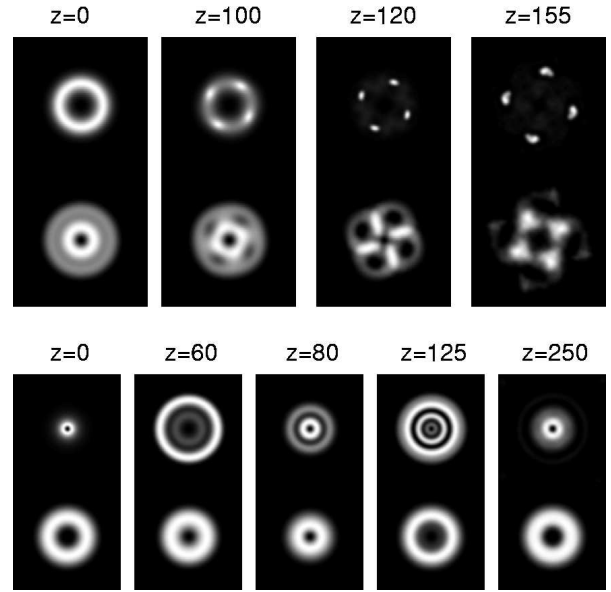}} \caption{\label{funny_prop}
Examples of the vortex instability
scenarios. Top: decay of type IV vortex solitons corresponding to
point G ($\mu=0.49$, $\beta=1.95$). Bottom: quasi-stable
propagation of the vortex with breathing components for type II
vector solitons (point H; $\mu=0.5$, $\beta=1.95$).}
\end{figure}

Shape of these solutions can be  explained qualitatively by a
simple analysis. In fact, from Eq.~(\ref{bounded_states_1}),
assuming $\mu$ small and the maximum amplitude of $v(r)$ much
larger than that of $u(r)$ (i.e. the parameter $\beta$ close to
the upper cutoff), we obtain the condition $\max[u^2/(1+\sigma
v^2)]>1$. Since $v$ has a larger maximum, the former condition
cannot be be satisfied unless the value of $v$ is small at the
position of the maximum of $u$, so that there still exists an
effective potential with a relative maximum higher than 1. That
requires that both fields have maxima shifted enough to each
other, explaining the shape of the solutions in region 5 of
Fig.~\ref{domain_first_mode}. When $\mu$ is larger, however, this
condition can be satisfied when the maxima of both solutions
almost coincide. Because the field $v$ becomes self-guided and the
shifting is not possible (due to the coupling between both
components), each field has to be localized in the region close to
the other. This explains the behavior of the solutions when $\mu$
grows and the solution crosses the boundaries between region 5 and
region 3, and then moves to region 1. On the other hand, if
$\beta$ is decreased, the component $v(r)$ becomes smaller being
guided by the component $u(r)$. This explains the behavior of the
solutions crossing the boundaries from region 5 into region 4 and
to region 2.

Existence of different kinds of double-vortex vector solitons
leads to multi-stability phenomena as well as more complicated
bifurcation diagrams. In Fig.~\ref{bifurcation_first_mode1} and
Fig.~\ref{bifurcation_first_mode2}, we present two examples of the
bifurcation diagrams for two different values of the coupling
parameter. In the first case, at $\mu=0.7$, only one solution for
each value of the propagation constant exists. At $\beta=1$, both
solutions of type I ($\beta>1$, region 1 in
Fig.~\ref{domain_first_mode}) and type II ($\beta <1$, region 2)
merge together. In fact, for $\beta=1$, it is deduced from
Eqs.~(\ref{stationary_vortex_induced_vaweguides}) that both
solutions $u(r)$ and $v(r)$ become identical.

For smaller values of $\mu$, however, the existence of different
kinds of vortex solutions generate a variety of branches in the
bifurcation diagram, as shown in
Fig.~\ref{bifurcation_first_mode2}. In this case, there exist both
the bifurcation points O$_1$ and O$_2$ and the critical points
O$_3$, O$_4$ and O$_5$, and three of the four kinds of solutions
exist for some values of $\beta$. For other values of the coupling
parameter $\mu$, different types of the bifurcation diagrams are
obtained, and they all show a change of the solutions when a
boundary between different existence domains is crossed.

In Fig.~\ref{stable_prop}, we show two examples of stable
propagation. The top rows present the evolution of the vector
soliton corresponding to point D in Fig.~\ref{domain_first_mode}.
For comparison, we show unstable propagation of the first
component alone, when the second vortex component is removed from
the input. In the latter case, the vortex decays after propagating
for some distance while the two component soliton remains
virtually stable for much longer propagation distance owing to the
coupling between both the components. In general, it is possible
to achieve a double-vortex vector soliton which remains stable for
an arbitrary distance provided we chose a state close enough to
the upper cutoff, where one of the components has larger amplitude
and the interaction between both components is strong. In the
bottom rows, we show another example of the stable propagation
corresponding to point E in Fig.~\ref{domain_first_mode}.

Different types of the double-vortex vector solitons demonstrate a
rich variety of the instability-induced scenarios of their
evolution. In Fig.~\ref{decaying_states} we present two examples
of the vortex evolution with two characteristic scenarios of the
vortex decay, producing either 2+4 (e.g. point F) or 2+2 (e.g.
point C) fundamental solitons. Finally, in Fig.~\ref{funny_prop}
we show two more complicated scenarios of the vortex instability
to illustrate a variety of the patterns that can be observed. In
the top row, where the input state corresponds to point G, the
soliton decays in a complex way displaying a sequence of symmetric
patterns. In the bottom row, where the input state corresponds to
point H, the double-vortex soliton propagates in a quasi-stable
way performing breathing radial oscillations. In all the cases
discussed here, the topological charges of both the components are
chosen with the same sign ($+1$). The case of the opposite charges
has also been studied, and the similar evolution scenarios have
been observed.


\section{Conclusions}

We have analyzed the existence and basic properties of the
two-component composite optical beams carrying an angular
momentum, the so-called vector vortex solitons. We have considered
two major types of such solitons that propagate in self-focusing
saturable nonlinear media and can be classified in the
low-intensity limit through the fundamental and first-order
localized modes guided by the main vortex beam. We have calculated
the existence domains of such vortex composite solitons and
studied numerically their stability to weak perturbations. In
particular, we have revealed a novel mechanism for stabilizing the
vortex azimuthal modulational instability by a co-propagating
guided mode of a large amplitude. We have demonstrated, for the
first time to our knowledge, the existence of bistable composite
double-vortex solitons as well as studied their
instability-induced dynamics. We believe that similar results can
be obtained for composite vortices in other types of nonlinear
models describing the mutual coupling between several fields, and
our results can be useful for other fields such as the physics of
the multi-species Bose-Einstein condensates of ultra-cold atoms.

\begin{acknowledgments}
This work was partially supported by the Australian Research
Council. J.R. Salgueiro acknowledges a postdoctoral fellowship of
the Secretar\'{\i}a de Estado de Educaci\'on y Universidades of
Spain supported by the European Social Fund, and he thanks
Nonlinear Physics Center at the Research School of Physical
Sciences and Engineering for a warm hospitality during his stay in
Canberra.
\end{acknowledgments}

\end{document}